\documentclass[10pt]{article}
\usepackage{graphicx}
\begin{document}
\title{Conserved excitation number and $U(1)$-symmetry operators for the anti-rotating (anti-Jaynes-Cummings) term of the Rabi Hamiltonian}
\author{Joseph Akeyo Omolo\\
Department of Physics, Maseno University, P.O. Private Bag, Maseno, Kenya\\
e-mail:~ ojakeyo04@yahoo.co.uk}
\date{12 November 2017}
\maketitle

\emph{Introduction}
The quantum Rabi model describes the dynamics of a quantized electromagnetic field mode interacting with a two-level atom generated by Hamiltonian [1-5]
$$H_R=\frac{1}{2}\hbar\omega\left(\hat{a}^\dagger\hat{a}+\hat{a}\hat{a}^\dagger\right)+\hbar\omega_0s_z+\hbar g(\hat{a}+\hat{a}^\dagger)(s_++s_-)  \eqno(1a)$$
where $\omega$ , $\hat{a}$ , $\hat{a}^\dagger$ are the quantized field mode angular frequency, annihilation and creation operators, while $\omega_0$ , $s_z$ , $s_+$ , $s_-$ are the atomic state transition angular frequency and operators. We have used $\sigma_x=s_-+s_+$ and expressed the free field mode Hamiltonian in the symmetrized normal and anti-normal order form $\frac{1}{2}\hbar\omega(\hat{a}^\dagger\hat{a}+\hat{a}\hat{a}^\dagger)$ for reasons which will become clear below.

Collecting the normal and anti-normal order terms in equation (1$a$), we express the Rabi Hamiltonian in the symmetrized form
$$H_R=\frac{1}{2}(~H+\overline{H}~) \eqno(1b)$$
where we have identified the normal order rotating component as the Jaynes-Cummings Hamiltonian $H$ obtained as
$$H=\hbar(~\omega\hat{a}^\dagger\hat{a}+\omega_0s_z+2g(\hat{a}s_++\hat{a}^\dagger s_-)~) \eqno(1c)$$
and the anti-normal order anti-rotating component as the anti-Jaynes-Cummings Hamiltonian $\overline{H}$ obtained as
$$\overline{H}=\hbar(~\omega\hat{a}\hat{a}^\dagger+\omega_0s_z+2g(\hat{a}s_-+\hat{a}^\dagger s_+)~) \eqno(1d)$$
We observe that the operator ordering principle which distinguishes the rotating (Jaynes-Cummings) and anti-rotating (anti-Jaynes-Cummings) components $H$ , $\overline{H}$ is not arbitrary, but has physical foundation. Noting that an electromagnetic field mode is composed of positive and negative frequency components [6], we provide the physical interpretation that the Jaynes-Cummings interaction represents the coupling of the atomic spin to the rotating positive frequency component of the field mode, such that the algebraic operations which generate the resulting \emph{red-sideband} state transitions are achieved through normal-operator-ordering, while the anti-Jaynes-Cummings interaction represents the coupling of the atomic spin to the anti-rotating negative frequency component of the field mode, such that the algebraic operations which generate the resulting \emph{blue-sideband} state transitions are achieved through anti-normal-operator-ordering. We note that blue-sideband effects arising from interactions involving negative frequency radiation have been observed in recent experiments [ 7 , 8 ].

In [2], the Jaynes-Cummings and ant-Jaynes-Cummings Hamiltonians $H$ , $\overline{H}$ have been characterized as the \emph{chiral} and \emph{anti-chiral} components, respectively, of the Rabi Hamiltonian. In this respect, we generalize the models of interaction between a single quantized field mode and a single two-level atom to include the asymmetric (anisotropic) Rabi models [3 , 4] by introducing a \emph{rotation-symmetry or chirality parameter $r$} taking values $-1\le r \le 1$ to express the Rabi Hamiltonian in equation (1$b$) in general symmetrization form
$$H_R=\frac{1}{2}(~(1+r)H+(1-r)\overline{H}~) \quad\quad ; \quad\quad -1\le r \le 1 \eqno(1e)$$
such that $r=1 , 0 , -1$ specifies that the Rabi Hamiltonian takes respectively the fully rotating (Jaynes-Cummings) , symmetric (isotropic) or fully anti-rotating (anti-Jaynes-Cummings) form, while for all other values $r\ne 1 , 0 , -1$ the Rabi Hamiltonian is asymmetric (anisotropic).

A major challenge, which has remained an outstanding problem in the Rabi model over the years, is that while the Jaynes-Cummings component has a conserved excitation number operator and is invariant under the corresponding $U(1)$-symmetry operation [1 , 3], a conserved excitation number and corresponding $U(1)$-symmetry operators for the anti-Jaynes-Cummings component have never been determined, leading to the general belief that the anti-Jaynes-Cummings interaction violates energy conservation principle. We address the long outstanding problem of excitation number and $U(1)$-symmetry operators of the anti-Jaynes-Cummings Hamiltonian in this letter.

\emph{Excitation number operators}
It follows from the physical interpretation given above that an excitation number operator in the Jaynes-Cummings interaction should be defined in normal-order form, while an excitation number operator in the anti-Jaynes-Cummings interaction should be defined in anti-normal-order form. Taking this operator ordering principle into account, we add and subtract an atomic spin normal order term $\hbar\omega s_+s_-$ in equation (1$c$) and anti-normal order term $\hbar\omega s_-s_+$ in equation (1$d$), then reorganize, noting $s_+s_-=\frac{1}{2}+s_z$ ~,~ $s_-s_+=\frac{1}{2}-s_z$, to obtain the Jaynes-Cummings Hamiltonian in the form
$$H=\hbar\omega(\hat{a}^\dagger\hat{a}+s_+s_-)+2\hbar g(\alpha s_z+\hat{a}s_++\hat{a}^\dagger s_-~)-\frac{1}{2}\hbar\omega \quad\quad ; \quad\quad  \alpha=\frac{\omega_0-\omega}{2g} \eqno(2a)$$
and the anti-Jaynes-Cummings Hamiltonian in the form
$$\overline{H}=\hbar\omega(\hat{a}\hat{a}^\dagger+s_-s_+)+2\hbar g(\overline{\alpha}s_z+\hat{a}s_-+\hat{a}^\dagger s_+~)-\frac{1}{2}\hbar\omega  \quad\quad ; \quad\quad \overline{\alpha}=\frac{\omega_0+\omega}{2g} \eqno(2b)$$
where we factored out $2g$ and introduced respective dimensionless frequency-detuning parameters $\alpha$ , $\overline{\alpha}$ defined as indicated.

In the Jaynes-Cummings Hamiltonian $H$, we identify the normally-ordered \emph{Jaynes-Cummings excitation number operator} $\hat{N}$, while in the anti-Jaynes-Cummings Hamiltonian $\overline{H}$, we identify the anti-normally ordered \emph{anti-Jaynes-Cummings excitation number operator} $\hat{\overline{N}}$, respectively defined by
$$\hat{N}=\hat{a}^\dagger\hat{a}+s_+s_- \quad\quad ; \quad\quad \hat{\overline{N}}=\hat{a}\hat{a}^\dagger+s_-s_+ \eqno(2c)$$
which we introduce in equations (2$a$) , (2$b$) as appropriate to express the Jaynes-Cummings and anti-Jaynes-Cummings Hamiltonians in the form
$$H=\hbar\omega\hat{N}+2\hbar g(~\alpha s_z+\hat{a}s_++\hat{a}^\dagger s_-~)-\frac{1}{2}\hbar\omega \quad\quad ; \quad\quad
\overline{H}=\hbar\omega\hat{\overline{N}}+2\hbar g(~\overline{\alpha}s_z+\hat{a}s_-+\hat{a}^\dagger s_+~)-\frac{1}{2}\hbar\omega \eqno(2d)$$
We observe that the Jaynes-Cummings excitation number operator $\hat{N}=\hat{a}^\dagger\hat{a}+s_+s_-$ is a standard conserved operator in the dynamics generated by the \emph{rotating component} of the Rabi Hamiltonian, while the anti-Jaynes-Cummings excitation number operator $\hat{\overline{N}}=\hat{a}\hat{a}^\dagger+s_-s_+$, which we establish here as a conserved operator in the dynamics generated by the \emph{anti-rotating component} of the Rabi Hamiltonian, is a new operator discovered and presented for the first time in the present letter. The discovery of the anti-Jaynes-Cummings excitation number operator, proof of its conservation and specification of the corresponding $U(1)$ and parity symmetry operators in the dynamics generated by the anti-Jaynes-Cummings Hamiltonian are the main results of this letter.

\emph{Proof of conservation : state transition operators}
Using standard atomic spin and field mode operator algebraic relations
$$[s_+ , s_-]=2s_z \ ; \quad [s_z , s_-]=-s_- \ ; \quad [s_z , s_+]=s_+ \ ; \quad s_+s_-=\frac{1}{2}+s_z \ ; \quad s_-s_+=\frac{1}{2}-s_z $$
$$[s_+s_- \ , \ s_+]=s_+ \ ; \quad [s_-s_+ \ , \ s_+]=-s_+ \ ; \quad [s_+s_- \ , \ s_-]=-s_- \ ; \quad [s_-s_+ \ , \ s_-]=s_-$$
$$\hat{a}\hat{a}^\dagger=\hat{a}^\dagger\hat{a}+1 \quad ; \quad [\hat{a}^\dagger\hat{a} \ , \ \hat{a}]=-\hat{a} \quad ; \quad
[\hat{a}^\dagger\hat{a} \ , \ \hat{a}^\dagger]=\hat{a}^\dagger \eqno(3a)$$
we easily prove that the excitation number operators $\hat{N}$ , $\hat{\overline{N}}$ in equation (2$c$) commute with the respective Jaynes-Cummings and anti-Jaynes-Cummings Hamiltonians $H$ , $\overline{H}$ in equation (2$d$) according to
$$[~\hat{N} \ , \ H~]=0 \quad\quad ; \quad\quad [~\hat{\overline{N}} \ , \ \overline{H}~]=0 \eqno(3b)$$
which proves the standard dynamical property that the excitation number operator $\hat{N}=\hat{a}^\dagger\hat{a}+s_+s_-$ is conserved in the dynamics generated by the Jaynes-Cummings Hamiltonian $H$ and the \emph{new} dynamical property that the excitation number operator $\hat{\overline{N}}=\hat{a}\hat{a}^\dagger+s_-s_+$ is conserved in the dynamics generated by the anti-Jaynes-Cummings Hamiltonian $\overline{H}$.

To make the proof even more transparent, we introduce two new conserved dynamical operators, namely, \emph{Jaynes-Cummings state transition operator} $\hat{A}$ and \emph{anti-Jaynes-Cummings state transition operator} $\hat{\overline{A}}$, respectively defined by
$$\hat{A}=\alpha s_z+\hat{a}s_++\hat{a}^\dagger s_- \quad\quad ; \quad\quad \hat{\overline{A}}=\overline{\alpha}s_z+\hat{a}s_-+\hat{a}^\dagger s_+ \eqno(3c)$$
which on squaring and applying standard atomic spin and field mode operator algebraic relations
$$s_z^2=\frac{1}{4} \ ; \quad s_-^2=s_+^2=0  \ ; \quad s_+s_-+s_-s_+=1 \ ; \quad s_zs_++s_+s_z=0 \ ; \quad s_zs_-+s_-s_z=0 \ ; \quad \hat{a}\hat{a}^\dagger=\hat{a}^\dagger\hat{a}+1 \eqno(3d)$$
provide the respective Jaynes-Cummings and anti-Jaynes-Cummings excitation number operators $\hat{N}$ , $\hat{\overline{N}}$ defined in equation (2$c$) in the form
$$\hat{A}^2=\hat{a}^\dagger\hat{a}+s_+s_-+\frac{1}{4}\alpha^2~=~\hat{N}+\frac{1}{4}\alpha^2 \quad\quad ; \quad\quad \hat{\overline{A}}^2=\hat{a}\hat{a}^\dagger+s_-s_++\frac{1}{4}~\overline{\alpha}^2-1~=~\hat{\overline{N}}+\frac{1}{4}~\overline{\alpha}^2-1 \eqno(3e)$$
Substituting $\hat{A}$ , $\hat{\overline{A}}$ from equation (3$c$) and $\hat{N}=\hat{A}^2-\frac{1}{4}\alpha^2$ , $\hat{\overline{N}}=\hat{\overline{A}}^2-\frac{1}{4}\overline{\alpha}^2+1$ from equation (3$e$) into equation (2$d$), we express the Jaynes-Cummings and anti-Jaynes-Cummings Hamiltonians in terms of the respective state transition operators in the form
$$H=\hbar(~\omega\hat{A}^2+2g\hat{A})-\frac{1}{4}\hbar\omega\alpha^2-\frac{1}{2}\hbar\omega \quad\quad ; \quad\quad
\overline{H}=\hbar(~\omega\hat{\overline{A}}^2+2g\hat{\overline{A}})-\frac{1}{4}\hbar\omega\overline{\alpha}^2 + \frac{1}{2}\hbar\omega \eqno(3f)$$
Using equations (3$e$) and (3$f$) easily confirms the commutation relations in equation (3$b$). In addition, it is easy to establish that the Jaynes-Cummings excitation number operator is not conserved in the dynamics generated by the anti-Jaynes-Cummings Hamiltonian $\overline{H}$ and likewise, the anti-Jaynes-Cummings excitation number operator is not conserved in the dynamics generated by the Jaynes-Cummings Hamiltonian $H$ according to the commutation relations
$$[~\hat{N}\ , \ \overline{H}~]\ne 0 \quad\quad ; \quad\quad [~\hat{\overline{N}}\ , \ H~]\ne 0 \eqno(3g)$$
Similarly, the state transition operators $\hat{A}$ , $\hat{\overline{A}}$ are conserved in the dynamics generated by the respective Hamiltonians $H$ , $\overline{H}$, but not in the dynamics generated by the other component Hamiltonian according to the commutation relations
$$[~\hat{A} \ , \ H~]=0 \quad\quad ; \quad\quad [~\hat{A} \ , \ \overline{H}~]\ne 0 \quad\quad ; \quad\quad
[~\hat{\overline{A}} \ , \ \overline{H}~]=0 \quad\quad ; \quad\quad [~\hat{\overline{A}} \ , \ H~]\ne 0\eqno(3h)$$
We have thus proved the desired conservation of the anti-Jaynes-Cummings excitation number operator and the state transition operators. We have established in another paper [9] that the Jaynes-Cummings state transition operator $\hat{A}$ generates \emph{red-sideband transitions} between \emph{polariton qubit states} arising in the rotating Jaynes-Cummings interaction $2\hbar g(\hat{a}^\dagger s_-+\hat{a}s_+)$, while the anti-Jaynes-Cummings state transition operator $\hat{\overline{A}}$ generates \emph{blue-sideband transitions} between \emph{anti-polariton qubit states} arising in the anti-rotating anti-Jaynes-Cummings interaction $2\hbar g(\hat{a}s_++\hat{a}^\dagger s_+)$. Eigenvectors and eigenvalues of the respective Jaynes-Cummings and anti-Jaynes-Cummings Hamiltonians $H$ , $\overline{H}$ in equation (3$f$), interpreted as polariton and anti-polariton qubit Hamiltonians, have been determined easily in [9]. The coupling of the atomic spin to the anti-rotating negative frequency component of the field mode leading to blue-sideband transitions accounts for the excitation number and energy conservation in the anti-Jaynes-Cummings interaction.

\emph{$U(1)$-symmetry operators}
The Jaynes-Cummings excitation number operator $\hat{N}=\hat{a}^\dagger\hat{a}+s_+s_-$ generates a free time evolution operator $U_0(t)$ obtained as
$$U_0(t)=e^{-i\omega t\hat{N}} \quad\quad\Rightarrow\quad\quad U_0^\dagger(t)=e^{i\omega\hat{N}t} \eqno(4a)$$
which provides field mode and atomic spin operator time evolution in the form
$$U_0^\dagger(t)\hat{a}U_0(t)=e^{-i\omega t}\hat{a} \ ; \quad U_0^\dagger(t)\hat{a}^\dagger U_0(t)=e^{i\omega t}\hat{a}^\dagger \ ; \quad U_0^\dagger(t)s_-U_0(t)=e^{-i\omega t}s_- \ ; \quad U_0^\dagger(t)s_+U_0(t)=e^{i\omega t}s_+ \eqno(4b)$$
The operator $U_0(t)$ thus generates symmetry operations on the Jaynes-Cummings and anti-Jaynes-Cummings Hamiltonians in equation (2$d$) in the form
$$U_0^\dagger(t)HU_0(t)=H \quad\quad ; \quad\quad U_0^\dagger(t)\overline{H}U_0(t)=\hbar\omega\hat{\overline{N}} +
2\hbar g(~\overline{\alpha}s_z+e^{-2i\omega t}\hat{a}s_-+e^{2i\omega t}\hat{a}^\dagger s_+~)-\frac{1}{2}\hbar\omega \eqno(4c)$$
which shows that the Jaynes-Cummings excitation number operator generated free time evolution operator $U_0(t)=e^{-i\omega t\hat{N}}$ is a $U(1)$-symmetry operator of the Jaynes-Cummings Hamiltonian $H$, but \emph{not} a symmetry operator of the anti-Jaynes-Cummings Hamiltonian $\overline{H}$.

On the other hand, the anti-Jaynes-Cummings excitation number operator $\hat{\overline{N}}=\hat{a}\hat{a}^\dagger+s_-s_+$ generates a free time evolution operator $\overline{U}_0(t)$ obtained as
$$\overline{U}_0(t)=e^{-i\omega t\hat{\overline{N}}} \quad\quad\Rightarrow\quad\quad \overline{U}_0^\dagger(t)=e^{i\omega\hat{\overline{N}}t} \eqno(4d)$$
which provides field mode and atomic spin operator time evolution in the form
$$\overline{U}_0^\dagger(t)\hat{a}\overline{U}_0(t)=e^{-i\omega t}\hat{a} \ ; \quad
\overline{U}_0^\dagger(t)\hat{a}^\dagger\overline{U}_0(t)=e^{i\omega t}\hat{a}^\dagger \ ; \quad
\overline{U}_0^\dagger(t)s_-\overline{U}_0(t)=e^{i\omega t}s_- \ ; \quad \overline{U}_0^\dagger(t)s_+\overline{U}_0(t)=e^{-i\omega t}s_+ \eqno(4e)$$
The operator $\overline{U}_0(t)$ thus generates symmetry operations on the Jaynes-Cummings and anti-Jaynes-Cummings Hamiltonians $H$ , $\overline{H}$ in equation (2$d$) in the form
$$\overline{U}_0^\dagger(t)\overline{H}\overline{U}_0(t)=\overline{H} \quad\quad ; \quad\quad \overline{U}_0^\dagger(t)H\overline{U}_0(t)=\hbar\omega\hat{N}+2\hbar g(~\alpha s_z+e^{-2i\omega t}\hat{a}s_++e^{2i\omega t}\hat{a}^\dagger s_-~) -\frac{1}{2}\hbar\omega \eqno(4f)$$
which shows that the anti-Jaynes-Cummings excitation number operator generated free time evolution operator
$\overline{U}_0(t)=e^{-i\omega t\hat{\overline{N}}}$ is a $U(1)$-symmetry operator of the anti-Jaynes-Cummings Hamiltonian $\overline{H}$, but \emph{not} a symmetry operator of the Jaynes-Cummings Hamiltonian $H$.

\emph{Parity-symmetry operator}
It follows from equations (4$c$) and (4$f$) that we can determine a \emph{common} symmetry operator of both Jaynes-Cummings and anti-Jaynes-Cummings Hamiltonians $H$ , $\overline{H}$ by imposing the free evolution symmetry condition
$$e^{-2i\omega t}=e^{2i\omega t}~=~1 \quad\quad\Rightarrow\quad\quad 2\omega t=2n\pi \quad ; \quad \omega t=n\pi \quad ; \quad n=1 , 2 , 3 , ...\eqno(5a)$$
where $n=0$ defines the identity operator. Substituting $\omega t=n\pi$ into equations (4$a$) , (4$d$), we obtain the common Jaynes-Cummings and anti-Jaynes-Cummings symmetry operator $\hat{\Pi}_n(\pi)$ in the form
$$\hat{\Pi}_n(\pi)=U_0(n\pi)~=~e^{-in\pi\hat{N}}~=~\overline{U}_0(n\pi)~=~e^{-in\pi\hat{\overline{N}}} \quad ; \quad n=1 , 2 , 3 , ...\eqno(5b)$$
which we express in the form
$$\hat{\Pi}_n(\pi)=(e^{-i\pi\hat{N}}~)^n~=~(e^{-i\pi\hat{\overline{N}}}~)^n~=~(\hat{\Pi})^n \eqno(5c)$$
from which we identify the standard Jaynes-Cummins and anti-Jaynes-Cummings parity-symmetry operator $\hat{\Pi}$ defined here by
$$\hat{\Pi}=e^{-i\pi\hat{N}}~=~e^{-i\pi\hat{\overline{N}}} \eqno(5d)$$
Substituting $\hat{N}=\hat{a}^\dagger\hat{a}+s_+s_-$ , $\hat{\overline{N}}=\hat{a}\hat{a}^\dagger+s_-s_+$ and using algebraic relations
$$\hat{a}\hat{a}^\dagger=\hat{a}^\dagger\hat{a}+1 \quad ; \quad\quad s_-s_+=s_+s_--2s_z \quad ; \quad\quad \hat{\overline{N}}=\hat{N}+2s_-s_+ \eqno(5e)$$
we obtain
$$e^{-i\pi\hat{\overline{N}}}=e^{-i\pi\hat{N}}e^{-2i\pi s_-s_+} \quad ; \quad e^{-2i\pi s_-s_+}=I \quad\Rightarrow\quad e^{-i\pi\hat{\overline{N}}}=e^{-i\pi\hat{N}} \eqno(5f)$$
which establishes the common Jaynes-Cummings and anti-Jaynes-Cummings parity-symmetry operator relation in equation (5$d$).

It is easy to establish that the Jaynes-Cummings and anti-Jaynes-Cummings parity-symmetry operator $\hat\Pi$ is a symmetry operator of the Rabi Hamiltonian $H_R=\frac{1}{2}(H+\overline{H})$ in equation (1$b$) according to the symmetry transformation operations
$$\hat{\Pi}^\dagger~H~\hat{\Pi}=H \quad\quad ; \quad\quad \hat{\Pi}^\dagger~\overline{H}~\hat{\Pi}=\overline{H} \quad\quad ; \quad\quad \hat{\Pi}^\dagger~H_R~\hat{\Pi}=H_R  \eqno(5g)$$
Finally, we observe that an important dynamical feature emerges from the Jaynes-Cummings-anti-Jaynes-Cummings common parity-symmetry relation in equation (5$d$). Substituting $\hat{N}=\hat{A}^2-\frac{1}{4}\alpha^2$ , $\hat{\overline{N}}=\hat{\overline{A}}^2-\frac{1}{4}\overline{\alpha}^2+1$ from equation (3$e$) into equation (5$d$) and reorganizing, we obtain the common parity-symmetry relation in the form
$$e^{-i\pi\hat{A}^2}~=~e^{-i\pi\hat{\overline{A}}^2}e^{i\pi(\frac{1}{4}\overline{\alpha}^2-\frac{1}{4}\alpha^2-1)}\eqno(6a)$$
which on using $\alpha=\frac{\omega_0-\omega}{2g}$ , $\overline{\alpha}=\frac{\omega_0+\omega}{2g}$ from equations (2$a$) , (2$b$) to evaluate
$$\frac{1}{4}\overline{\alpha}^2-\frac{1}{4}\alpha^2=\frac{\omega_0\omega}{4g^2}~=~\beta^2 \eqno(6b)$$
takes the form
$$e^{-i\pi\hat{A}^2}~=~e^{-i\pi\hat{\overline{A}}^{~2}}e^{i\pi(\beta^2-1)} \quad\quad ; \quad\quad \beta^2=\frac{\omega_0\omega}{4g^2} \eqno(6c)$$
which suggests that there exists a \emph{critical coupling constant} $g_c$ at which the global phase factor $e^{i\pi(\beta^2-1)}$ equals unity obtained as
$$g=g_c \quad ; \quad e^{i\pi(\beta_c^2-1)}=1 : \quad\quad \beta_c^2=\frac{\omega_0\omega}{4g_c^2}~=~1 \quad\quad\Rightarrow\quad\quad g_c=\frac{1}{2}\sqrt{\omega_0\omega} \eqno(6d)$$
giving common parity-symmetry relation at the critical coupling $g_c$ in the form
$$g=g_c \ : \quad \alpha_c=\overline{\alpha}_c=\frac{\omega_0-\omega}{2g_c} \quad\quad ; \quad\quad \beta_c^2=\frac{\omega_0\omega}{4g_c^2} \quad\quad ; \quad\quad \hat{\Pi}_c=e^{-i\pi\hat{A}_c^2}~=~e^{-i\pi\hat{\overline{A}}_c^{~2}} \eqno(6e)$$
We identify $g_c=\frac{1}{2}\sqrt{\omega_0\omega}$ to be exactly the critical coupling constant at which the Rabi interaction undergoes quantum phase transition as determined in a recent study [5]. It follows that parity-symmetry breaking may occur at a quantum phase transition. We have presented quantum phase transition phenomena in the Rabi and the more general Dicke models in another paper.

\emph{Conclusion}
We have applied operator-ordering as the fundamental algebraic property to determine the conserved excitation number and $U(1)$-symmetry operators for the rotating (Jaynes-Cummings) and anti-rotating (anti-Jaynes-Cummings) components of the Rabi Hamiltonian. The specification of the anti-Jaynes-Cummings excitation number operator means that the eigenvalue spectrum of the anti-Jaynes-Cummings Hamiltonian can now be determined explicitly. The Rabi Hamiltonian is thus composed of two algebraically complete Jaynes-Cummings and anti-Jaynes-Cummings components, each specified by its characteristic excitation number, state transitions, $U(1)$-symmetry and red or blue sideband eigenvalue spectrum. We have determined the parity-symmetry operator as the common symmetry operator for both Jaynes-Cummings and anti-Jaynes-Cummings components, leading to the standard algebraic property that parity operator is a symmetry operator of the Rabi Hamiltonian. The parity-symmetry may break at a critical coupling constant $g_c$ where quantum phase transition occurs.

\emph{Acknowledgement}
I thank Maseno University for providing facilities and a conducive work environment during the preparation of the manuscript.


\begin{thebibliography}{30}
\bibitem{Bra} D~Braak 2011 On the Integrability of the Rabi Model, Phys.Rev.Lett.{\bf 107}, 100401 ; arXiv:1103.2461 v2 [quant-ph]
\bibitem{Cun} B~C~da Cunha, M~C~de Almeida and A~R~de Queiroz 2016 On the existence of monodromies for the Rabi model, J.Phys.A: Math.Theor.{\bf 49}, 194002
\bibitem{Xie} Q~Xie, H~Zhong, M~T~Batchelor and C~Lee 2016 The quantum Rabi model: solutions and dynamics, arXiv:1609.00434 v2 [quant-ph]
\bibitem{Xie} Q~Xie, et al 2014 Anisotropic Rabi Model, Phys.Rev.{\bf X 4}, 021046 ; arXiv:1401.5865v2[quant-ph]
\bibitem{Hwa} M~J~Hwang, R~Puebla and M~B~Plenio 2015 Quantum phase transition and universal dynamics in the Rabi model, Phys.Rev.Lett.~{\bf 115}, 180404 ; arXiv: 1503.03090 [quant-ph]
\bibitem{Bor} M~Born and E~Wolf, 1999 \emph{Principles of Optics}, Cambridge University Press, Cambridge, England
\bibitem{Rub} E~Rubino, et al 2012 Negative-Frequency Resonant Radiation, Phys.Rev.Lett.~{\bf 108}, 253901
\bibitem{McL} J~McLenaghan and F~Konig 2014 Few-cycle fiber pulse compression and evolution of negative resonant radiation, New J. Phys..~{\bf 16}, 063017
\bibitem{Omo} J~A~Omolo 2018 Polariton and anti-polariton qubits in the Rabi model, Preprint-ResearchGate, DOI:10.13140/RG.2.2.11833.67683




\end{thebibliography}
\end{document}